# Variations in magnetic properties of nanostructured nickel

Paramita Kar Choudhury*,[1] S. Banerjee,[2] S. Ramaprabhu,[3] K. P. Ramesh,[1] and Reghu Menon[1]

[1]*Department of Physics, Indian Institute of Science, Bangalore 560012, India*
[2]*Saha Institute of Nuclear Physics, 1/AF, Bidhannagar, Kolkata 700 064, India*
[3]*Department of Physics, Indian Institute of Technology Madras, Chennai-600 036, India*

The magnetic properties of carbon nanotube encapsulated nickel nanowires (C.E. nanowires of diameter ~ 10 nm), and its comparison to other forms of Ni are carried out in this work. The saturation magnetization ($M_s$) and coercivity ($H_c$) for C.E. nanowires are 1.0 emu/g and 230 Oe. The temperature dependence of coercivity follows $T^{0.77}$ dependence indicating a superparamagnetic behavior. The field-cooled and zero-field-cooled plots indicate that the blocking temperature ($T_B$) ~ 300 K. These altered magnetic properties of C.E. nanowires are mainly due to the nanoscale confinement effect from carbon nanotube encapsulation. The shape and magnetic environment enhance the total magnetic anisotropy of C.E. nanowires by a factor of four.

* Electronic address: s_paramita@physics.iisc.ernet.in

Fax: +91-80-2360 2602

Tel: +91-80-2293 2859

## 1. INTRODUCTION

Magnetic nanostructures have received increasing interest because of their potential application in magnetic recording media, sensors, and storage systems, and are fundamentally important as model systems to investigate nanomagnetism which is significantly different from magnetic properties of bulk material.[1-4]

Earlier studies in the magnetic nanosystems have shown that the basic magnetic properties can vary significantly as a function of both size and shape.[5] For e.g. the saturation magnetization ($M_s$ ~ 57.8 emu/g ) and coercivity ($H_c$ ~ 0.7 Oe) values in bulk nickel are shifted to 2.5 emu/g and 193.6 Oe, respectively, as the particle size is reduced to 2-3 nm.[6] The results have shown that coercivity undergoes a change as the particle size is reduced to nanoscale, reaches a maximum for a critical diameter, and then tends to decrease.[7] This is attributed to the variation from multi to single-domain regime where dipolar interactions, magnetic anisotropy energy as well as the packing density play crucial roles.[7-9] Although magnetic properties as a function of size and shape have been investigated, a systematic study for a particular system by taking into account the aspect ratio and surrounding environments is yet to be reported.[5,8] Since it is known that nanomagnetism is quite sensitive to the intrinsic (size and shape of the particles) and extrinsic (coatings, encapsulation, etc.) contributions, a comparative investigation is required to separate out these factors.

This paper addresses the basic issue of how magnetism in nickel is transformed as a function of size and form in different environments. Towards this, a typical sample of carbon nanotube encapsulated (C.E.) Ni nanowires has been studied, and the effect of encapsulation on magnetic properties has been determined. In this work, the magnetic properties of C.E. nanowires are measured and compared to that of reported works on bulk,[10] nanorods[11] and template embedded nanowires[12] (T.E. nanowires). The experimental data show that in C.E. nanowires: $H_c$ ~ 230 Oe, blocking temperature ~ 300 K and shape anisotropic constant is $2 \times 10^5$ erg / $cm^3$, which are quite different from other forms of Ni. Furthermore the shielding of nanowires with diamagnetic carbon nanotube layer provides an effective barrier against oxidation, which is suitable for high temperature magnetic applications in storage devices, sensors, etc.

## 2. EXPERIMENTAL DETAILS

Polycrystalline nanowires of nickel are grown inside multiwall carbon nanotubes (MWNT) by thermal chemical vapor deposition (CVD) using $LaNi_2$ as catalyst.[13] The TEM images in Fig. 1 reveal that nickel nanowire of diameter ~ 10 nm is encapsulated within MWNT of outer diameter ~ 30 nm. In a typical collection of these C.E. nanowires, the nanowires are randomly oriented. The magnetic measurements were carried out in a superconducting quantum interference device (SQUID) magnetometer [MPMS – 7 (Quantum Design)]. For zero field cooled (ZFC) measurement, the sample was cooled down to 5 K in absence of magnetic field, and the data were taken while warming to 300 K in presence of a magnetic field of 100 Oe. Field cooled (FC) data were taken while warming the sample after cooling it to 5 K at 100 Oe. The hysteresis data were taken at 5, 100 and 300 K, up to 7 Tesla. The data of bulk Ni, of particle size of several micrometers, is taken as a reference.[10] Moreover, magnetic properties of randomly oriented Ni nanorods, synthesized in presence of hexadecylamine (HDA) is chosen for comparison.[11]

Data for Ni nanowires embedded in porous alumina membrane [named as template embedded (T.E.) nanowires], are also used for this comparative study.[12] These samples were selected in such a way that the diameter of each of them (except the bulk) is in the range of 4-30 nm, so that the magnetic properties are not significantly altered due to size dependent contributions.

## 3. RESULTS AND DISCUSSION

The hysteresis loop of C.E. nanowires has been compared to that for bulk and T.E. nanowires at 300 K, and for nanorods at 2 K, as shown in Fig.2. The value of $M_S$ for nanorods (60 emu/g at 2 K), close to that of bulk Ni, reduces drastically for C.E. nanowires (0.8 emu/g). The reduction in net magnetization is mainly due to the random orientation of the C.E. nanowires. The $H_C$ values vary considerably for nanorods (275 Oe at 2 K), T.E. nanowires along longitudinal (750 Oe) and transverse (180 Oe) directions, and in CE nanowires (230 Oe). Also, it is good to compare the saturation field ($H_S$) values for nanorods (~ 4000 Oe), T.E. nanowires along longitudinal (~ 200 Oe) and transverse (~ 2700 Oe) directions, and in CE nanowires (~ 15,000 Oe). It is interesting to note that $M_S$, $H_C$ and $H_S$ differ significantly from sample to sample, although the basic nanocrystalline structure prevails in all these Ni samples. It is well known that both number and size of domains play a major role in magnetic properties.[7] The values of $H_C$, for nanoscale samples are nearly two orders of magnitude larger than that of bulk, and it decreases with increasing size due to the transition from single domain to multidomain regime. As the size of nanostructure increases, coercivity decreases due to the increase in the number of domain walls. Whitney *et al* have reported that $H_C$ varies from 650 to 400

Oe, as the size increases from 50 to 100 nm.[9] However, nanoscale samples below a certain critical diameter (typically less than 30 nm), having a single magnetic domain, are considerably affected by interparticle dipolar interaction and surface anisotropy energy, besides the thermal fluctuations. Zheng *et al* have observed that $H_c$ increases by a factor of two (550 to 950 Oe) as the diameter of Ni wires increases from 10 to 20 nm indicating that the magnetic properties are rather sensitive at nanoscale.[8]

In this work, the diameter (4-30 nm) of nanoscale samples are chosen in such a way that domain dependent roles in the nanomagnetic properties are minimized; since all the samples, except bulk, are expected to be in single-domain regime. It is interesting to note that the magnetic environment is quite different in each of these cases. The values of $H_c$ in longitudinal (transverse) direction are high (low) for T.E. nanowires, and rather low for C.E. nanowires, which is mainly due to the contributions from magnetic environments. The nanorods are unshielded since they are tethered by HAD, which is not expected to alter the magnetization at low fields. The T.E. nanowires are embedded within alumina template so that the field can easily penetrate the nanowires in longitudinal measurements, leading to large magnetization. This explains the large value for $H_c$ in longitudinal direction, and the low value of $H_s$ to reach saturation magnetization. However, in transverse direction, diamagnetic shielding due to alumina membrane offers resistance to magnetic field, this result in a rather low value of $H_c$, as well as a higher value of $H_S$. Since C.E. nanowires are fully enclosed within diamagnetic carbon nanotubes, the $H_c$ and $M_S$ values are less than that of T.E. nanowires due to the screening of field; and subsequently increase the value of $H_S$. Although in Fig. 2, the $M/M_s$ vs. $H$ plots for C.E. nanowires and nanorods look similar, this is mainly due to the fact that the data for latter is at 2 K, with a different temperature dependence of coercivity.[11]

The hysteresis loops for C.E. nanowires at 5, 100 and 300 K are shown in Fig. 3 (a). The coercivity values plotted as a function of temperature is shown in Fig. 3(b). The temperature dependence of coercivity for single domain particles is given by the relation: [14,15]

$$H_C = H_{C0}\,[1-(T/T_B)^k\,] \qquad (1)$$

where $H_{C0}$ is zero temperature coercivity, and $T_B$ is blocking temperature. The exponent $k$ is 0.5 for an assembly of aligned particles and 0.77 for randomly oriented particles. It is well known that at nanoscale confinement, where the size is of the order of 5-50 nm, the thermal activation energy overcomes cohesive energy of fluctuating magnetic domains, above a certain blocking temperature ($T_B$), and this provides adequate energy for the alignment of particle moments in an applied magnetic field. As a result, the hysteretic behavior is suppressed and the system behaves as a strong paramagnet or a superparamagnet at $T > T_B$.[7] Our data in Fig. 3(b) roughly follows the $T^{0.77}$ behavior and suggests the value of $T_B \geq 300$ K for unaligned C.E. nanowires. The deviation from the ideal behavior can be explained by understanding the different mechanisms responsible for magnetization reversal in nanoparticles and nanowires. Although $T^{0.5}$ dependence of coercivity has been reported in nanoparticles,[14,16] the mechanism for the temperature dependence of coercivity in nanowires is not yet clearly understood. Although the coercivity value for the nanorods (Fig. 2) is high (275 Oe) at 2 K, it should be considerably reduced (following Eq. 1) at 300 K, since the value of $T_B \sim 100$ K.

Furthermore, the FC-ZFC plot in Fig. 3 (c) shows a bifurcation at 300 K, which provides further evidence for the suggested superparamagnetic behavior. The temperature dependence of FC curve, from 5 - 300 K, resembles that of a paramagnet while the onset of superparamagnetic regime is indicated by the bifurcation at 300 K, as observed in both heating and cooling cycle data. The reported value of $T_B$ (298 K) for Ni nanoparticles of diameter 30-50 nm is very similar to that for the C.E. nanowires. [17] Bulk Ni shows ferromagnetic behavior, whereas various nanostructures of Ni (e.g. nanoparticles, particle clusters and nanocrystals, size 5-40 nm) exhibit superparamagnetism.[18,19] For nanowires several microns long but diameters < 50 nm, superparamagnetism has also been reported.[20,21] The values of $T_B$ for nanoscale samples are shown in Table I, indicating that larger the particle size, higher the value of $T_B$.

For $T \geq T_B$, the magnetization vs. $H/T$ curves scale to Langevin type behavior:[14]

$$\frac{M_H}{M_s} = L(a) = \coth(a) - \frac{1}{a} \qquad (2)$$

where $a=(\mu_{eff} H)/ k_B T$. The effective moment is given by $\mu_{eff} = M_s <V>$, where $<V>$ is particle volume and $k_B$ is Boltzmann constant. The magnetization data for C.E. nanowires at 300 K fit well to Eq. 2, as shown in Fig. 3 (d). The data have been corrected to remove diamagnetic contribution of the carbon nanotube encapsulation. As determined from the fit: $\mu_{eff} \sim 1.1 \times 10^{-16}$ erg / Oe = $1.2 \times 10^4$ $\mu_B$, where $\mu_B$ is Bohr Magneton. Such high value of magnetic moment has been reported for similar nanosystems that show superparamagnetic behavior.[6] Accumulation of large number of uncompensated spins on the surface of magnetically isolated nanowires can give rise to such high magnetic moment which in turn can explain the superparamagnetism in encapsulated nanowires.

Since the aspect ratio of these nano samples differs considerably, the anisotropic contribution to their magnetic properties has been evaluated in detail. The effective anisotropy in a magnetic particle ($K_p$) results from both magnetocrystalline anisotropy ($K_u$) and shape anisotropy ($K_s$) contributions: $K_p = K_u + K_s$. Nevertheless for spherical nanoparticles, only magnetocrystalline anisotropy contributes to $K_p$. In case of Ni and other transition metal based magnetic materials, the magnetocrystalline anisotropy is not high; instead the shape anisotropy plays an important role in controlling the magnetic properties. The shape anisotropy contribution in nanowires can be expressed as[22]

$$K_s = \frac{(Hs_{ll} - Hs_{\perp})(M_s)}{2} \qquad (3)$$

where $(Hs_{ll} - Hs_{\perp}) = 2\pi M_s$ is the effective saturation field in parallel and perpendicular directions of aligned nanowires. The value of $K_s$ for T.E nanowires is calculated by using Eq. 3, as shown in Table I. In C.E. nanowires, the nanowires being randomly oriented, the direction dependent measurement cannot be carried out. As a result the above equation cannot be used and the shape anisotropy constant ($K_s$) has been determined by an alternate method.

For a particle exhibiting superparamagnetism, the effective anisotropy ($K_p$) is determined by the relation [11]

$$T_B = K_p V / 25\, k_B \qquad (4)$$

where $V$ is the particle volume. It is to be noted that, the contribution due to dipolar interactions[23] has not been taken into account in this equation since the particle volume in our system is quite large (length ~ 100 nm and diameter ~ 10 nm). Moreover the diamagnetic encapsulation due to carbon-walls increases the interparticle separation; as a result the dipolar interaction is considerably weakened in the C.E. nanowires. By using the known values of $V = 7850\ nm^3$ and $T_B$ ~ 300 K, the value of $K_p = 1.5 \times 10^5$ erg/cm$^3$, as obtained from Eq. 4. By subtracting the value of magnetocrystalline anisotropy constant for nickel[24] [$K_u = - 0.5 \times 10^5$ erg/cm$^3$] from the total anisotropy ($K_p$), the value of shape anisotropy is determined as $K_s \sim 2 \times 10^5$ erg/cm$^3$. Comparing the values of $K_u$ and $K_s$ shows that the shape of C.E. nanowires increases the total anisotropy ($K_p$) by a factor of four. This value for C.E. nanowire is quite close to other reported samples as shown in Table I. It is interesting to note that since $K_u$ is an intrinsic property that cannot be altered, but the value of $K_s$ can be tuned to a large extent by controlling the aspect ratio, particularly by varying the diameter of nanotube/rod/wire; and this can be achieved by optimizing the sample preparation techniques. Thus by tuning the shape of nanoscale structure and its environment, it is possible to control the total anisotropy of the material, which can be used for various applications that demand directionality in magnetic properties.

## 4. CONCLUSIONS

To summarize, the TEM images show that the nickel nanowires (C.E.) are quite well encapsulated within multi-wall carbon nanotubes. The diamagnetic shielding of carbon nanotube acts as an effective barrier against oxidation that can be very useful for high temperature magnetic applications. The magnetization of C.E. nanowires are compared with respect to bulk, nanorods and template embedded (T.E.) nanowires indicating that size, shape and environment play a crucial role in determining the values of $M_s$, $H_c$ and $H_s$. In C.E. nanowires, $H_c$ follows $T^{0.77}$ dependence, and the bifurcation in FC-ZFC plot shows superparamagnetic behavior with $T_B$ ~ 300 K. Such superparamagnetic behavior in long nanowires with confined diameter can be ascribed to the large value of effective magnetic moment obtained from Langevin fit to the magnetization data. Also observed that the value of total anisotropy constant ($K_p$) of the system can be controlled by varying the shape of the nanostrucutres.

## References

[1] K. Liu, K. Nagodawithana, P. C. Searson, and C. L. Chien, Phys. Rev. B 51**,** 7381 (1995).

[2] D. J. Sellmeyer, M. Yu, R. A. Thomas, Y. Liu, and R. D. Kirby, Phys. Low-Dimens. Semicond. Struct. 1/2**,** 155 (1998).

[3] L. Zhentao, H. Chao, Y. Chang, and Q. Jieshan, J. Nanosci. Nanotechnol. 9, 7473 (2009).

[4] D. Meneses-Rodríguez, E. Muñoz-Sandoval, G. Ramírez-Manzanares, D. Ramírez-González, S. Díaz-Castañon, J. C. Faloh-Gandarilla, A. Morelos-Gómez, F. López-Urías, M. Terrones, J. Nanosci. Nanotechnol. 10, 5576 (2010).

[5] I. M. L. Billas, A. Chatelain, Walt A. de Heer, Science 265, 1682 (1994).

[6] M. A. Khadar, V. Biju, and A. Inoue, Mater. Res. Bull. 38, 1341 (2003).

[7] B. D. Cullity, and C. D. Graham, Introduction to Magnetic Materials, IEEE Press, Wiley Publication.

[8] M. Zheng, L. Menon, H. Zeng, Y. Liu, S. Bandyopadhyay, R. D. Kirby, and D. J. Sellmyer, Phys. Rev. B 62, 12282 (2000).

[9] T. M. Whitney, J. S. Jiang, P. C. Searson, and C. L. Chien, Science 261, 1316 (1993).

[10] L. Daroczi, D.L.Beke, G.Posgay, and M. Kis-Varga, Nanostructured Materials 6, 981 (1995).

[11] N. Cordente, M. Respaud, F. Senocq, M. -J. Casanove, C. Amiens, and B. Chaudret, Nanolett. 1, 565 (2001).

[12] M. Va´zquez, K. Pirota, M. Herna´ndez-Ve´ lez, V. M. Prida, D. Navas, R. Sanz, F. Batalla´n, and J. Vela´zquez, J. Appl. Phys. 95, 6642 (2004).

[13] A. Leela Mohana Reddy, M. M. Shaijumon and S. Ramaprabhu, Nanotechnology 17, 5299 (2006).

[14] E. M. Brunsman, R. Sutton, E. Bortz, S. Kirkpatrick, K. Midelfort, J. Williams, P. Smith, M. E. McHenry, S. A. Majetich, J. O. Artman, M. De Graef, and S. W. Staley, J Appl. Phys. 75, 5882 (1994).

[15] H. Pfeiffer, and W. Schiippel, Phys. Status Solidi A 119, 259 (1990).

[16] X.-C. Sun, and X. -L. Dong, Mater. Res. Bull. 37, 991 (2002).

[17] Y. Chen, D. -L. Peng, D. Lin and X. Luo, Nanotechnology 18, 505703 (2007).

[18] L. Yue, R. Sabiryanov, E. M. Kirkpatrick, and D. L. Leslie-Pelecky, Phys. Rev. B 62, 8969 (2000).

[19] C. P. Bean, and J. D. Livingston, J. Appl. Phys. 30, 120 S (1959).

[20] C.-C. Chen, Y.-, J. Hsu, Y.-, F. Lin, and S.-, Y. Lu, J. Phys. Chem. C 112, 17964 (2008).

[21] L. Zhang, and Y. Zhang, J. Mag. and Mag. Mater. 321, L15 (2009).

[22] K. M. Razeeb, F. M. F. Rhen, and S. Roy, J Appl. Phys. 105, 083922 (2009).

[23] R. Das, A. Gupta, D. Kumar, S. H. Oh, S. J. Pennycook, and A. F. Hebard, J. Phys: Condens. Matter 20, 385213 (2008).

[24] M. B. Stearns. Magnetic Properties of Metals: 3d, 4d, and 5d Elements, Alloys and Compounds, Springer Publisher Vol. 19a (1986).

## Figure Captions

Fig.1. TEM and HRTEM (inset) image of multiwall carbon nanotube encapsulated (C.E.) Ni nanowires.

Fig. 2. $M/M_S$ vs. H plot at 300 K for C.E. nanowires as compared to other forms of Ni: bulk and template embedded (T.E.) nanowires at 300 K, and the data for nanorods at 2 K.

Fig. 3. (a) Hysteresis plots for C.E. nanowires at different temperatures (b) Temperature dependence of coercivity, solid line is guide to eyes (c) FC-ZFC plots at 100 Oe, showing $T_B$ ~ 300 K (d) M vs. H/T plot at 300 K, solid line is Langevin fit to Eq.2.

Table I: Comparison of magnetic properties [size, $H_C$, blocking temperature ($T_B$), magnetocrystalline ($K_u$) and shape ($K_s$) anisotropy constants] for various forms of Ni in this work.

| Sample | Size (diameter × length) *nm* | $H_C$ (Oe) | $T_B$ (K) | $K_u$ (erg/cm$^3$) | $K_s$ (erg/cm$^3$) |
|---|---|---|---|---|---|
| Bulk [Ref. 10] | $\geq 10^3$ | 0.7 | -- | $-0.5 \times 10^5$ (at 300 K) | -- |
| Nanorods (random) [Ref. 11 ] | $4 \times 15$ | 275 (at 2 K) | 100 | $-7 \times 10^5$ (at $T_B$) | $7.7 \times 10^5$ |
| Template embedded (T.E.) nanowires (aligned) [Ref. 12] | $30 \times 100$ | Longitudinal 750 Transverse 180 | -- | $-0.5 \times 10^5$ (at 300 K) | $5 \times 10^5$ |
| Carbon nanotube encapsulated (C.E.) nanowires (random) | $10 \times 100$ | 230 | 300 | $-0.5 \times 10^5$ (at $T_B$) | $2 \times 10^5$ |

Figures

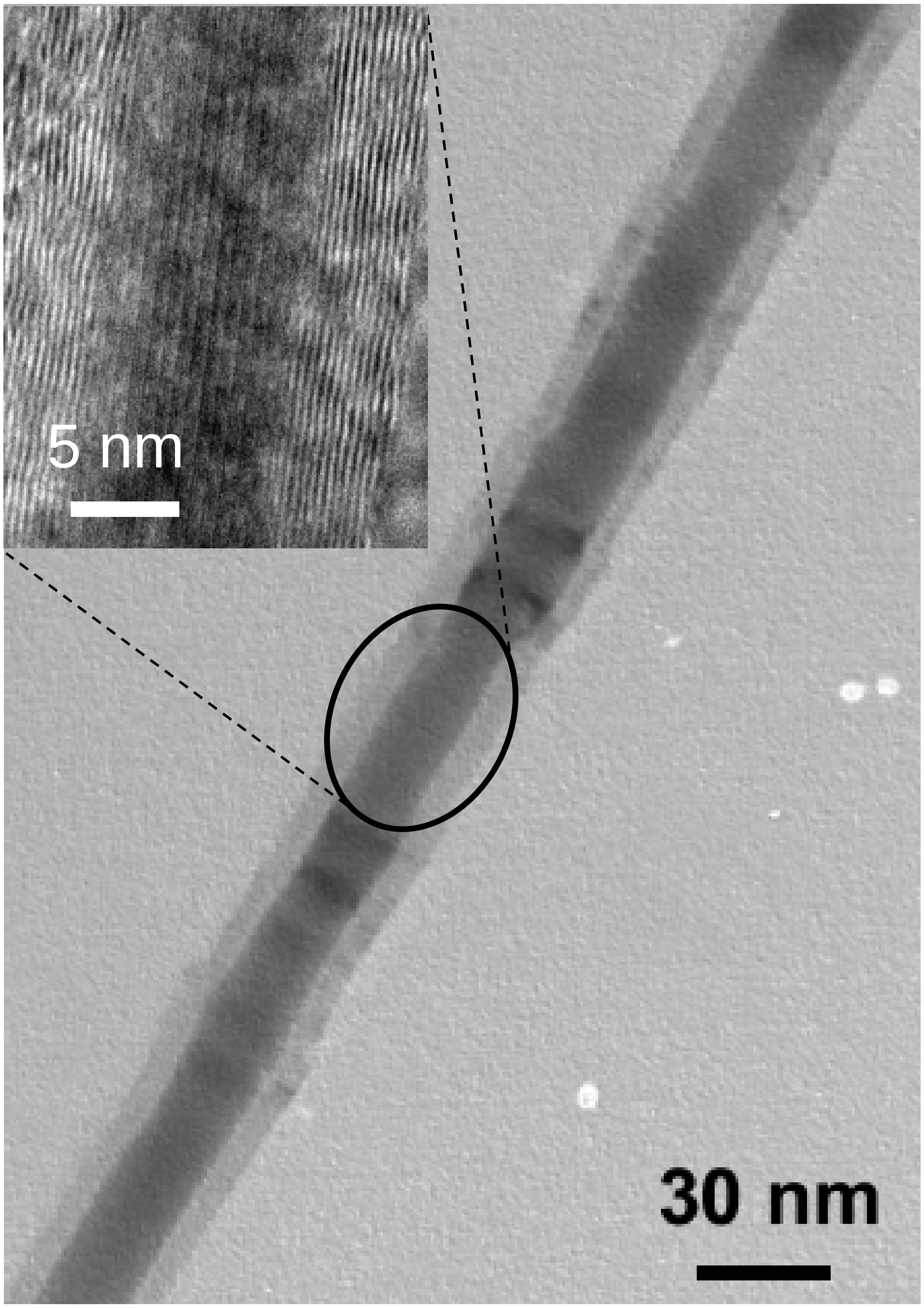

Fig 1. TEM and HRTEM (inset) image of carbon encapsulated (C.E.) Ni nanowires.

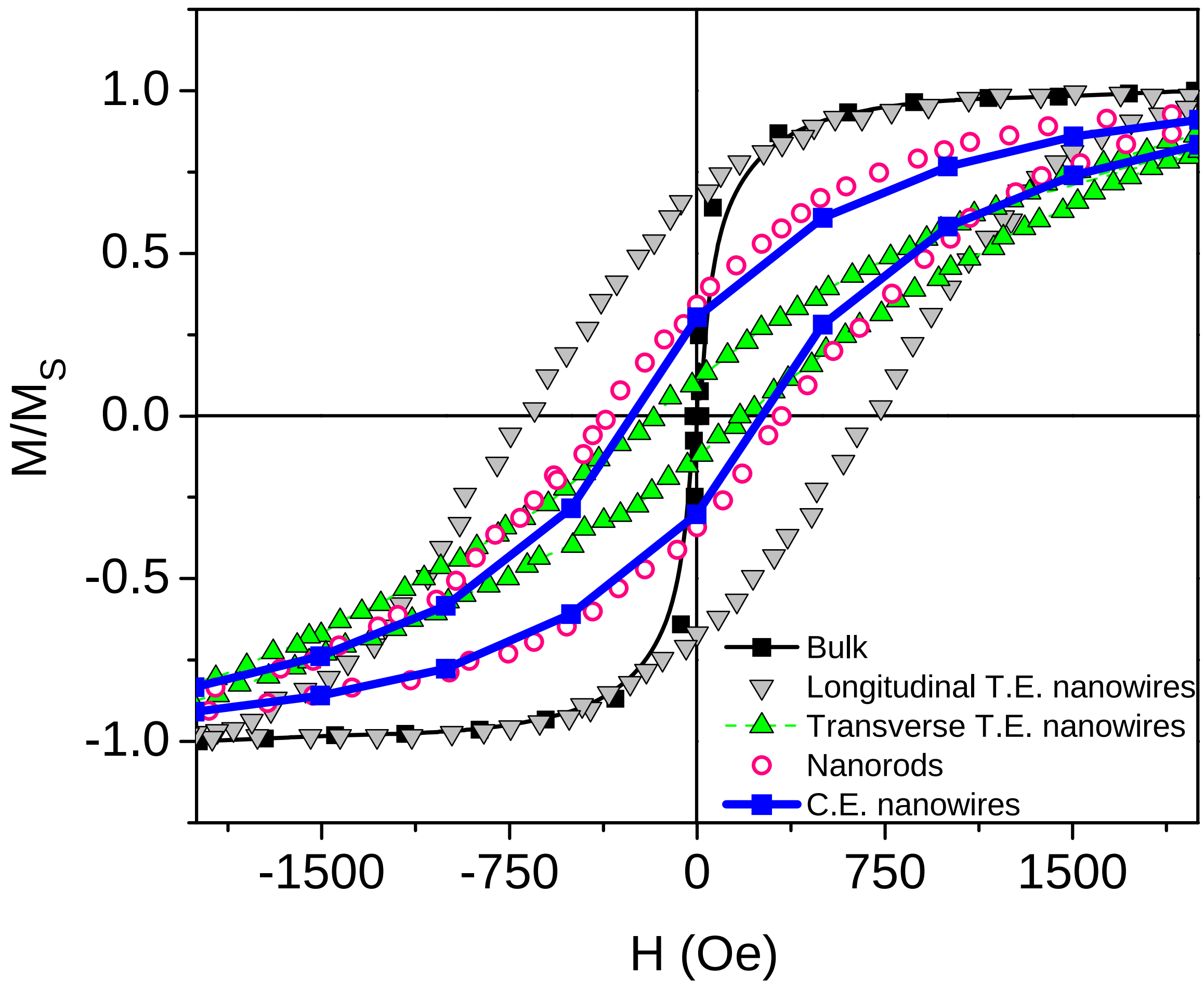

Fig. 2. M/M$_S$ vs. H plot at 300 K for C.E. nanowires as compared to other forms of Ni: bulk and template embedded (T.E.) nanowires at 300 K, and the data for nanorods at 2 K.

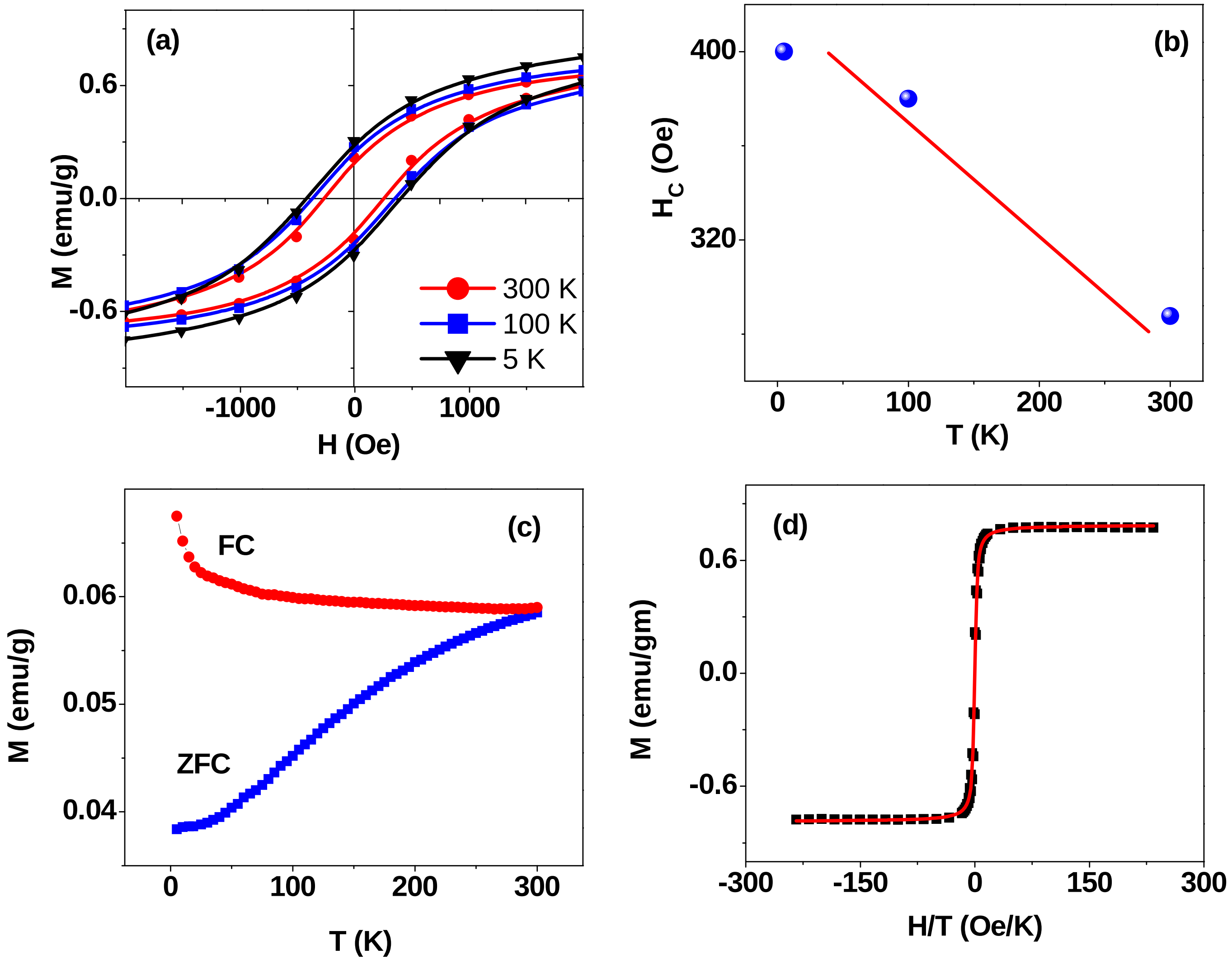

Fig. 3. (a) Hysteresis plots for C.E. nanowires at different temperatures (b) Temperature dependence of coercivity, solid line is a guide to eyes (c) FC-ZFC plots at 100 Oe showing $T_B$ ~ 300 K (d) M vs. H/T plot at 300 K, solid line is Langevin fit to Eq.2.